\documentclass[a4paper,12pt]{article}

\usepackage{amsmath,amssymb,graphicx,amsthm,lineno}

\usepackage[none]{hyphenat}
\usepackage{authblk}

\usepackage[utf8]{inputenc}
\usepackage[english]{babel}

\usepackage{verbatim}

\usepackage{color}
\usepackage{soul}

\usepackage[none]{hyphenat}
\usepackage{authblk}

\newcommand*{\ov}[1]{\overline{\raisebox{0pt}[1.2\height]{$#1$}}}

\usepackage{mdframed}

\title{Closed-form Second Solutions to the Coulombic Schr\"odinger Equation}

\author{William C. Parke}

\affil{Department of Physics\\ The George Washington University\\ Washington, D.C.}

\date{}

\begin{document}

\maketitle

\begin{abstract}
\sloppy
The regular solutions to the Sch\"ordinger equation in the case of an electron experiencing a Coulomb 
force are well known.  Being that the radial part of the differential equation 
to be solved is second order in derivatives, it will have two independent solutions, with the second irregular solutions 
being ill-behaved at the origin and unbound at infinity. For this reason, second 
solutions are dropped for bound electrons. However, these second solutions are
still of academic interest for several reasons. One reason is to help
devise schemes to control numerical contamination by second solutions to more
general second-order differential equations when attempting to
calculate the first solutions through recursion relations. Another is to study the
analytic behavior of the electron Coulomb wave functions as the electron energy and angular momentum are
each extended into the complex plane, important in investigations of bound-state poles
and of Regge trajectories. In addition, toy problems having a finite
radial region with a Coulombic interaction, with interior and exterior regions non Coulombic,
require both regular and irregular solutions to match solutions across the boundaries.
In this presentation, exact and closed form second irregular solutions are derived
for the Coulomb bound states, using a hither-to unnoticed trick in the Nikiforov-Uvorov
method.
\end{abstract} 

\section{Introduction}

In the following, we will focus on the second (irregular) solutions to the
Schr\"odinger equation for hydrogen-like atoms when the electron is in a
a steady state. As a second-order differential
equation, the general solution before imposing boundary conditions
is a linear combination of two independent solutions.  However, in
this case, the second solution produces
electron probabilities that diverge at
the origin when the orbital angular momentum quantum number $\ell$ is not zero, 
is not a solution near the origin when $\ell=0$. The irregular solution also diverges with distance
from the origin, and so the electron would be unbound. Therefore,
the second solution is usually dropped from the general solution. (See
Messiah \cite[p\,350-352]{Messiah}, for more details of this argument.)

Even so, a
second solution has utility.  It may be employed when the electron potential energy is
Coulombic only within a shell region around the nucleus. In that case,
the shell-region solution in general has both first and second solutions
added in order to match the wave function and its derivative on
the boundaries of the shell. Also, knowing the
second solution behavior is useful in formulating how to cure the second
solution contamination of numerical calculations of first solutions of 
more general second-order differential equations.
Thirdly, having the closed-form second solutions at hand helps in the study of the
analytic behavior of Coulomb scattered-wave amplitudes as the energy $E$ of
the scattered electron is extended into the complex plane, and 
as the electron angular momentum quantum $\ell$ takes on complex
values, such as in Regge pole analysis. (See \cite{gasp} and \cite{take}.)

\section{Review of the first solution}

The steady-state radial part $R(r)$ of the Schr\"odinger equation (SE) wave
function $\psi(r,\theta,\phi)=R(r)Y_{\ell}^{m_{\ell}}(\theta,\phi)$ for a lone electron
interacting with a nucleus carrying a charge $Z\left|e\right|$ 
satisfies 
\begin{equation}
\frac{\hbar ^{2}}{2\mu}\,\,\left(-\frac{1}{r^{2}}\frac{d }{d  r} r^{2}\frac{%
d }{d  r}R\left(r\right) +\ell (\ell+1) \frac{1}{r^{2}}\right) -
\left( E+\frac{Ze^{2}}{4\pi \epsilon _{0}r} 
\right)R(r) =0  \,\, .
\end{equation}
Here, $\mu$ is the reduced nucleus-electron mass, $E$ the total energy of
the system, $Y_{\ell}^{m_{\ell}}$ is a
spherical harmonic, an eigenstate of $\vec{L}^2$ and $L_z$, with
eigen\-values $\ell(\ell+1)\hbar^2$ and $m_{\ell}\hbar$, respectively,
where $\vec{L}\,$ is the angular momentum operator 
of the nucleus-electron system, assuming its center-of-mass is at rest.

For bound states, $E<0$. Let 
\begin{equation}
\kappa =\frac{\sqrt{-2\mu E}}{\hbar }\,,
\end{equation}
making 
\begin{equation}
r^{2}\frac{d  ^{2}R\left( r\right) }{d  r^{2}}+2r\frac{d 
R\left( r\right) }{d  r}+\left( -\kappa ^{2}r^{2}+\frac{2\mu e^{2}Z}{%
4\pi \epsilon _{0}\hbar ^{2}}r-\ell(\ell+1)\right) R\left( r\right) =0\,.
\end{equation}
For simplifying appearances and easing manipulations, define 
\begin{equation}
\kappa _{0}=\frac{\mu e^{2}Z}{4\pi \epsilon _{0}\hbar ^{2}}=\frac{\mu }{m_{e}%
}Z\frac{1}{a}\,,
\end{equation}
($a$ being the Bohr radius) so that the radial SE becomes  
\begin{equation}
r^{2}\frac{d  ^{2}R\left( r\right) }{d  r^{2}}+2r\frac{d 
R\left( r\right) }{d  r}+\left( -\kappa ^{2}r^{2}+2\kappa
_{o}r-\ell(\ell+1)\right) R=0\,.
\end{equation}
For large $r,$ $R\rightarrow ce^{-\kappa r}$. (For now, we drop the $%
e^{\kappa r}$ solution as we are considering first solutions for bound states.) Extracting the
asymptotic part, we define the function $u(r)$ by 
\begin{equation}
R(r)=e^{-\kappa r}u(r)
\end{equation}
giving 
\begin{equation}
r^{2}\frac{d  ^{2}u(r)}{d  r^{2}}+2r(1-\kappa r)\frac{d 
u(r)}{d  r}+\left( -\ell(\ell+1)+2\left( \kappa _{0}-\kappa \right) r\right)
u(r)=0
\end{equation}

Now, we will follow Nikiforov and Uvarov \cite{NU} to analyze the solutions of the differential
equations which take the form 
\begin{equation}
\sigma _{1}u^{\prime \prime }+\tau _{1}u^{\prime }+\frac{\sigma _{2}}{\sigma
_{1}}u=0 \, ,
\end{equation}
where the $\sigma$ are polynomials in $r$ of degree no greater than two, and
the $\tau$ are polynomials of degree no greater than one. Here, primes are
used to indicate differentiation with respect to the implied independent
variable.

They put this differential equation into `standard' hypergeometric form 
\begin{equation}
\sigma _{1}y^{\prime \prime }+\tau _{2}y^{\prime }+\lambda y=0 \ 
\end{equation}
by using a special choice of transformation 
\begin{equation}
u=\phi y \, .
\end{equation}
For convenience, we will let 
\begin{equation}
\phi \left( r\right) =e^{\varphi \left( r\right) } \, .
\end{equation}
Then {\footnotesize 
\begin{equation}
\frac{d  ^{2}y\left( r\right) }{d  r^{2}}+\left( 2\frac{d 
\varphi \left( r\right) }{d  r}+\frac{ \tau _{1}}{\sigma _{1}}\right) 
\frac{d  y\left( r\right) }{d  r}+\left( \frac{d 
^{2}\varphi \left( r\right) }{d  r^{2}}+\left( \frac{d  \varphi
\left( r\right) }{d  r}\right) ^{2}+\frac{\tau _{1}}{\sigma _{1}}\frac{%
d  \varphi \left( r\right) }{d  r}+\frac{\sigma _{2}}{\sigma
_{1}^{2}}\right) y\left( r\right) =0 \, .
\end{equation}
}

To gain simplicity, they try to select the function $\phi$ so that the
coefficient of the derivative term, $\left( 2\frac{d  \varphi \left(
r\right) }{d  r}+\frac{\tau _{1}} {\sigma_{1}}\right) $, has the form $%
\tau _{2}/\sigma _{1},$ where $\tau _{2}$ has degree no greater than one.
Assume this can be done.

Let 
\begin{equation}
\tau _{2}=2\sigma _{1}\frac{d  \varphi \left( r\right) }{d  r}%
+\tau _{1}
\end{equation}
which means 
\begin{equation}
\sigma _{1}\frac{d  \varphi \left( r\right) }{d  r}=\frac{1}{2}%
\left( \tau _{2}-\tau _{1}\right) \,
\end{equation}
is also of degree no more than one.

In terms of $\tau_2$, the coefficient of $y(r)$ is
\begin{equation}
\frac{d  ^{2}\varphi \left( r\right) }{d  r^{2}}+\left( \frac{%
d  \varphi \left( r\right) }{d  r}\right) ^{2}+\frac{\tau _{1}}{%
\sigma _{1}}\frac{d  \varphi \left( r\right) }{d  r}+\frac{%
\sigma _{2}}{\sigma _{1}^{2}} \,\, ,
\end{equation}
which becomes 
\begin{equation}
\frac{1}{2\sigma _{1}}\left( \tau _{2}^{\prime }-\tau _{1}^{\prime }\right)
-\left( \tau _{2}-\tau _{1}\right) \frac{1}{2\sigma _{1}^{2}}\sigma
_{1}^{\prime }+\left( \frac{1}{2\sigma _{1}}\left( \tau _{2}-\tau
_{1}\right) \right) ^{2}+\frac{\tau _{1}}{2\sigma _{1}^{2}}\left( \tau
_{2}-\tau _{1}\right) +\frac{\sigma _{2}}{\sigma _{1}^{2}} \, .
\end{equation}
Define 
\begin{equation}
\sigma _{3}=\frac{\sigma _{1}}{2}\left( \tau _{2}^{\prime }-\tau
_{1}^{\prime }\right) -\frac{1}{2}\sigma _{1}^{\prime }\left( \tau _{2}-\tau
_{1}\right) +\frac{1}{4}\left( \tau _{2}^{2}-\tau _{1}^{2}\right) +\sigma
_{2} \, .
\end{equation}
Evidently, $\sigma _{3}$ is of degree no greater than two. The differential
equation for $y$ becomes 
\begin{equation}
\frac{d  ^{2}y\left( r\right) }{d  r^{2}}+\frac{\tau _{2}}{%
\sigma _{1}}\,\frac{d  y\left( r\right) }{d  r}+\frac{\sigma _{3}}{%
\sigma _{1}^{2}} \, y\left( r\right) =0 \,\, .
\end{equation}

To take the wanted hypergeometric form, the coefficient of $y$ should be
proportional to $1/\sigma _{1},$ i.e. 
\begin{equation}
\sigma _{3}=\lambda \sigma _{1}
\end{equation}
for all $r.$ This is possible because there are three relations and three
unknows: $\left( \tau _{20},\tau _{21},\lambda \right) .$ Explicitly, we want 
\begin{equation}
\lambda \sigma _{1}=\frac{1}{2}\sigma _{1}\left( \tau _{2}^{\prime }-\tau
_{1}^{\prime }\right) -\frac{1}{2}\sigma _{1}^{\prime }\left( \tau _{2}-\tau
_{1}\right) +\frac{1}{4}\left( \tau _{2}^{2}-\tau _{1}^{2}\right) +\sigma
_{2} \,\, ,
\end{equation}
a quadratic in the variable $r$, to hold for all $r.$

In the Coulomb case, the equation 
\begin{equation}
\sigma _{1}u^{\prime \prime }+\tau _{1}u^{\prime }+\frac{\sigma _{2}}{\sigma
_{1}}u=0
\end{equation}
reads 
\begin{equation}
r\frac{d  ^{2}u\left( r\right) }{d  r^{2}}+2\left(1-\kappa
r\right) \frac{d  u\left( r\right) }{d  r}+\frac{-\ell(\ell+1)+2\left(
\kappa _{0}-\kappa \right) r}{r} u\left( r\right)=0 \,\, ,
\end{equation}
so 
\begin{eqnarray}
\sigma _{1} &=&r \, , \\
\tau _{1} &=&2\left( 1-\kappa r\right) \, , \\
\sigma _{2} &=&-\ell(\ell+1)+2\left( \kappa _{0}-\kappa \right) r \,\, .
\end{eqnarray}

Transforming\ the differential equation to standard hypergeometric form will
make 
\begin{equation}
\lambda \sigma _{1}=\frac{1}{2}\sigma _{1}\left( \tau _{2}^{\prime }-\tau
_{1}^{\prime }\right) -\frac{1}{2}\sigma _{1}^{\prime }\left( \tau _{2}-\tau
_{1}\right) +\frac{1}{4}\left( \tau _{2}^{2}-\tau _{1}^{2}\right) +\sigma
_{2}
\end{equation}
or, in terms of the coefficients in the polynomials, 
\begin{equation}
\lambda r= \left( -\frac{1}{2}\tau _{20}+\frac{1}{4}\tau _{20}^{2}-l\left(
l+1\right) \right) +\left( \frac{1}{2}\tau _{20}\tau _{21}+2\kappa
_{0}\right) r+\left( -\kappa ^{2}+\frac{1}{4}\tau _{21}^{2}\right) r^{2} \, .
\end{equation}

Thus, the two coefficients in $\tau _{2}=\tau _{20}+\tau _{21}r$ must
satisfy 
\begin{eqnarray}
\tau _{20} &=&2\ell+2,-2\ell\,, \\
\tau _{21} &=&-2\kappa \,\,,\,2\kappa \,\, ,
\end{eqnarray}
and the `eigenvalue' $\lambda $ will be 
\begin{eqnarray}
\lambda &=&\frac{1}{2}\tau _{20}\tau _{21}+2\kappa _{0} \\
&=&2\kappa _{0}-2\left( \ell+1\right) \kappa \, .
\end{eqnarray}

Nikiforov and Uvarov observed that to be able to get eigenfunction solutions having a non-negative
index (see equation (\ref{eq-lambda})), the polynomial $\tau _{2}$ must have a
negative derivative, and a zero, somewhere in $\left[ 0,\infty \right] ,$ so
we take 
\begin{eqnarray}
\tau _{20} &=&2\ell+2 \\
\tau _{21} &=&-2\kappa \,\, ,
\end{eqnarray}
i.e. 
\begin{equation}
\tau _{2}=2\left( \ell+1-\kappa r\right) \,\,.
\end{equation}

Now from 
\begin{equation}
\tau _{2}=2\sigma _{1}\frac{d  \varphi \left( r\right) }{d  r}%
+\tau _{1}
\end{equation}
\begin{eqnarray}
\frac{d  \varphi \left( r\right) }{d  r} &=&\frac{1}{2\sigma _{1}%
}\left( \tau _{2}-\tau _{1}\right) \\
&=&\frac{1}{2r}\left( 2\left( \ell+1-\kappa r\right) -2\left( 1-\kappa r\right)
\right) \\
&=&\frac{\ell}{r} \\
\varphi &=&\ell\ln r\,\,,
\end{eqnarray}
so we get 
\begin{equation}
\phi =r^{\ell}\,\,.
\end{equation}
From 
\begin{equation}
\sigma _{3}=\left( \frac{\sigma _{1}}{2}\left( \tau _{2}^{\prime }-\tau
_{1}^{\prime }\right) -\frac{1}{2}\sigma _{1}^{\prime }\left( \tau _{2}-\tau
_{1}\right) +\frac{1}{4}\left( \tau _{2}^{2}-\tau _{1}^{2}\right) +\sigma
_{2}\right)
\end{equation}
we have 
\begin{equation}
\sigma _{3}=-\frac{1}{2}+2\left( \kappa _{0}-\left( \ell+1\right) \kappa r
\right) \,\, .
\end{equation}

Eigenvalues for the number $\lambda $\ are determined by 
\begin{equation}
\lambda +n_{r}\tau _{2}^{\prime }+\frac{1}{2}n_{r}\left( n_{r}-1\right)
\sigma _{1}^{\prime \prime }=0 \,\, ,
\label{eq-lambda}
\end{equation}
where $n_r$ is a non-negative integer (later identified as the number of
radial nodes in the solution $R(r)$). Inserting the Coulomb case, 
\begin{equation}
2\kappa_{0}-2\left( \ell+1\right) \kappa -2\kappa n_{r} =0 \,\, .
\end{equation}
Thus, the $\kappa$ are quantized according to
\begin{equation}
\kappa =\kappa _{0}\frac{1}{n_{r}+\ell+1} \,\, .
\end{equation}

Using the definition of $\kappa$, we see that for the bound-state solution
to exist, the energy $E$ of the electron must be quantized according to 
\begin{eqnarray}
E &=&-\frac{\hbar ^{2}}{2m}\frac{1}{4}\kappa _{0}^{2}\frac{1}{\left(
n_{r}+\ell+1\right) ^{2}} \\
&=&-\frac{\hbar ^{2}}{2m}\left( \frac{mZ}{\hbar ^{2}}\right) ^{2}\left( 
\frac{e^{2}}{4\pi \epsilon _{0}}\right) ^{2}\frac{1}{\left( n_{r}+\ell+1\right)
^{2}} \\
&=&-\frac{mZ^{2}}{2\hbar ^{2}}\left( \frac{e^{2}}{4\pi \epsilon _{0}}\right)
^{2}\frac{1}{\left( n_{r}+\ell+1\right) ^{2}} \,\, ,
\end{eqnarray}
which are the energy levels first derived by Bohr.
Comparing with the Bohr formula, we can identify the positive integer 
\begin{equation}
n\equiv n_r+l+1
\end{equation}
as the Bohr quantum number (also, these days, called the 'principle' quantum
number).

The transformed differential equation becomes 
\begin{equation}
\sigma _{1}y^{\prime \prime }+\tau _{2}y^{\prime }+\lambda y=0 \,\, .
\end{equation}
In the Coulomb case, 
\begin{equation}
ry^{\prime \prime }+2\left( \ell+1-\kappa r\right) y^{\prime }+2\left( \kappa
_{0}-\left( \ell+1\right) \kappa \right) y=0 \,\, .
\end{equation}

Compare to the differential equation for the associated Laguerre
polynomials, $L_{1}(\ov{n},m,x)\equiv L_{\overline{n}}^{m}(x)$, which is 
\begin{equation}
xL^{\prime \prime }+\left( m+1-x\right) L^{\prime }+\ov{n}L=0\,\,.
\label{eq-laguerre}
\end{equation}
The comparison suggests we let 
\begin{equation}
x=2\kappa r\,\,,
\end{equation}
so 
\begin{equation}
x\frac{d  ^{2}}{d  x^{2}}y+\left( 2l+1+1-x\right) \frac{d  
}{d  x}y+\left( \frac{\kappa _{0}}{\kappa }-\left( l+1\right) \right)
y=0\,\,.
\end{equation}

From the above, we also know that quantization of the electron bound states
gives $\kappa=\kappa_n$, where the quantized $\kappa_n$ are fixed by 
\begin{equation}
\frac{\kappa _{0}}{\kappa_n }=n_{r}+\ell+1 \,\, ,
\end{equation}
so 
\begin{equation}
x\frac{d  ^{2}}{d  x^{2}}y+\left( 2\ell+1+1-x\right) \frac{d  
}{d  x}y+n_{r}y=0 \,\, .
\end{equation}

Thus, the 'regular' ('first') solutions for the differential equation for $%
y(x)$ are 
\begin{equation}
y_{1}(n,\ell,x)=L_{1}(n-\ell-1,2\ell+1,2\kappa _{n}r)\equiv L_{n_{r}}^{2\ell+1}(2\kappa
_{n}r)\,\,,
\end{equation}
and the radial SE `regular' solution is (up to a constant factor)
\begin{equation}
R_{1}(n,\ell,r)=(2\kappa _{n}r)^{\ell}e^{-\kappa _{n}r}L_{1}(n-\ell-1,2\ell+1,2\kappa
_{n}r)\,\,.
\end{equation}
where
\begin{equation}
L_{1}\left( n_r,m,x\right) =\sum_{j=0}^{n_r}\left( -1\right) ^{j}\left( \frac{%
\left( n_r+m\right) !}{\left( n_r-j\right) !(m+j)!}\right) \frac{1}{j!}x^{j} \,\, .
\end{equation}

\section{The second solution}

Consider the differential equation (\ref{eq-laguerre}) expressed as: 
\begin{equation}
xy^{\prime\prime}+ (m+1-x)y^{\prime}+\ov{n}y = 0  \label{dech}
\end{equation}
in which $m$ is a positive integer and $\ov{n}$ is a non-negative integer. (In the Coulomb
case, $\ov{n}=n_r$,\, $m=2l+1$.) 

As we have seen, equation \eqref{dech} has polynomial solutions proportional to the
associated Laguerre polynomials $L_{\overline{n}}^{m}(x)$. These polynomials are expressible in terms of the confluent
hypergeometric function as
\begin{equation}
L_{\overline{n}}^{m}\left( x\right) =\frac{\left( m+1\right) _{\ov{n}}}{\ov{n}!}%
\,\,_{1}F_{1}(-\ov{n},m+1,x)
\label{eq-L1}
\end{equation}
in which
\begin{equation}
\,_{1}\!F_{1}(-\ov{n},m+1,x)=\sum_{k=0}^{\overline{n}}\frac{(-\ov{n})_{k}}{%
(m+1)_{k}k!}x^{k},  \label{Phi}
\end{equation}
where $(-\ov{n})_{k}$ and $(m+1)_{k}$ are Pochhammer symbols, defined by 
$(a)_k\equiv a(a+1)\cdots (a+k-1)=\Gamma(a+k)/\Gamma(a)$, with $(a)_0= 1$.

Following Nikiforov and Uvarov, if we know a first solution $\Phi_1(\ov{n},m,x)$ to
the hypergeometric equation, a second linearly independent solution  
is given by the extended Cauchy integral:
\begin{equation}
\Phi_2(\ov{n},m,x) = \frac{1}{\rho(x)}\int_{0}^{\infty}\frac{\rho(s)\Phi_1(\ov{n},m,s)}{s-x%
}\,ds  \label{cipsi}
\end{equation}
in which the weight function $\rho(x)$ = $e^{-x}x^m$ is, for the
differential equation \eqref{dech}, a solution of the equation $%
(x\rho(x))^{\prime}= (m-x)\rho(x)$. The integral is taken
with principle value near the pole of the the integrand.

We now write \eqref{cipsi} in the form 
\begin{eqnarray}
\Phi_2 (\ov{n},m,x)&=&x^{-m}e^{x}\int_{0}^{\infty }\frac{e^{-s}}{s-x}\left[ s^{m}\Phi_1
(\ov{n},m,s)
-x^{m}\Phi_1 (\ov{n},m,x)\,\right] \,ds\notag\\
&+&\ \Phi_1 (\ov{n},m,x)\int_{-x}^{\infty }%
\frac{e^{-s}}{s}ds\ .  \label{Psi-sol0}
\end{eqnarray}
The great advantage ('trick') in this separation is that the first integral
no longer has a pole at $s=x$ AND the exponential factor $e^{-s}$ has
been taken out of the bracketed expression.

Inserting the  hypergeometric solution \eqref{Phi}, the first of
the two integrals in Eq.\thinspace \eqref{Psi-sol0} is 
\begin{samepage}
\begin{eqnarray}
&& \sum_{k=0}^{\overline{n}}\frac{(-\ov{n})_{k}}{(m+1)_{k}k!}\int_{0}^{\infty}\frac{e^{-s}}{(s-x)}\big[s^{m+k} - x^{m+k}\big]ds\notag \\
 &&= \sum_{k=0}^{\overline{n}}\frac{(-\ov{n})_{k}}{(m+1)_{k}k!}\sum_{j=0}^{m+k-1}(m+k-1-j)!\,x^{j}
\end{eqnarray}
\end{samepage}
while the last term of Eq.\thinspace \eqref{Psi-sol0} contains the
'standard' exponential-integral function 
\begin{equation}
 \mbox{Ei}\left( 1,-x\right) \equiv \int_{-x}^{\infty }\frac{e^{-s}}{s}\,ds\,\, 
\notag
\end{equation}
times the first solution.
There results 
\begin{equation}
\Phi_2 (\ov{n},m,x)=\frac{m!}{(\ov{n}+m)!}\,P_{2}(\ov{n},m,x)\frac{e^{x}}{x^{m}}+\,_1F_1\left(
-\ov{n},m+1,x\right) \mbox{Ei}(1,-x) \,\, , \label{Psi} 
\end{equation}
where the polynomials $P_{2}\left( \ov{n},m,x\right) $ are
\begin{eqnarray}
&&P_2\left( \ov{n},m,x\right) =\frac{(\ov{n}+m)!}{m!}\sum_{k=0}^{\overline{n}}\sum_{j=0}^{m+k-1}%
\frac{(-\ov{n})_{k}(m+k-1-j)!}{(m+1)_{k}\ k!}x^{j} \,\, .  \label{eq-P2}
\end{eqnarray}

The equation \eqref{Psi} constitutes an explicit
closed-form second solution to the confluent hypergeometric differential
equation in the degenerate case.

The normalization of the polynomial $P_2(\ov{n},m,x)$ has been chosen to
make the coefficient of $x^{\overline{n}+m-1}$ be $(-1)^{\overline{n}}$. It then turns out that all
the coefficients of the powers of $x$ are integers.

The polynomials $P_{2}(\ov{n},m,x)$ can be written as a sum of two
terms, the first with only positive coefficients and powers of $x$ up to $x^{m-1}$, and
a second with sign-oscillating terms with powers $x^m$ up to $x^{\overline{n}+m-1}$:
\begin{eqnarray}
&&P_{2}\left( \ov{n},m,x\right) =\sum_{p=0}^{m-1}\frac{\left( \ov{n}+p\right) !\left(
m-p-1\right) !}{p!}x^{p}\notag\\
&&-\,x^{m}\sum_{p=0}^{\overline{n}-1}\left[ \sum_{k=0}^{\overline{n}-p-1}%
\frac{\ov{n}!}{\left( \ov{n}-k-p-1\right) !}\frac{(\ov{n}+m)!}{\left( m+k+p+1\right) !}%
\frac{(-1)^{k}k!}{\left( k+p+1\right) !}\right] \left( -x\right) ^{p}. \notag\\
\end{eqnarray}
As shown by Parke and Maximon \cite{Parke-Max}, the bracketed coefficient in the second sum can
be simplified, so that
\begin{eqnarray}
&&P_{2}\left( \ov{n},m,x\right) =\sum_{p=0}^{m-1}\frac{\left( \ov{n}+p\right) !\left(
m-p-1\right) !}{p!}x^{p}\notag\\
&-&x^{m}\sum_{p=0}^{\overline{n}-1}\frac{(\ov{n}+m)!}{(m+p)!}%
\left(\sum_{k=m+p+1}^{\overline{n}+m}\frac{1}{k}\,(\ov{n}+m+1-k)_{p}\,\right)\,\frac{(-x)^{p}}{p!}
\,\, \notag\\
\end{eqnarray}
or
{\footnotesize
\begin{eqnarray}
&&P_{2}\left( \ov{n},m,x\right) =\sum_{p=0}^{m-1}\frac{\left( \ov{n}+p\right) !\left(
m-p-1\right) !}{p!}x^{p}\notag\\
&-&\frac{(\ov{n}+m)!}{m!}\,x^{m}\left( \sum_{k=m+1}^{\overline{n}+m}%
\frac{1}{k}+\sum_{p=1}^{\overline{n}-1}\frac{m!}{(m+p)!}\frac{\left( -x\right) ^{p}}{p!}%
\sum_{k=m+p+1}^{\overline{n}+m}\frac{1}{k}\prod_{j=1}^{p}(\ov{n}+m+j-k)\right)\,\, . \notag\\
\label{eq-Psimple}
\end{eqnarray}
}
(The value $\ov{n}=0$ is special, since the sum has an upper limit smaller than the
lower limit.  It is consistent to take such a sum as zero. Similarly, a product of this
type is taken as one.)

\section{Second radial solution}

From the above analysis, our Coulomb radial second solution is proportional to
\begin{equation}
\Psi (n_r,m,x)=\exp{(-x/2)}x^{-(m-1)/2}\Phi_2(n_r,m,x) \,\, ,
\label{eq-Psi}
\end{equation}
where $n_r=n-\ell-1,\,m=2\ell+1,$, and $x=2\kappa_n r$.

We will define the arbitrary coefficient which can be placed as a factor in
front of the solution $\Psi(n_r,m,x)$ to make the $Ei$ term (after using equation (\ref{Psi})) be
simply $\exp{(-x/2)}x^{-\ell}L_1(n_r,m,x)\,\,\mbox{Ei}(1,-x)$, where, from equation (\ref{eq-L1}),
\begin{equation*}
L_{1}\left( n_r,m,x\right) =\frac{(m+n_{r})!}{m!n_{r}!}\,_{1}F_{1}\left(
-n_{r},m+1,x\right) \, .
\end{equation*}

This means that our second (`irregular') solutions $R_2(n,\ell,r)$ to the steady-state (bound
electron) radial Schr\"{o}dinger equation can be written as

\vspace{0.2cm}

\begin{mdframed}

\begin{align}
R_{2}(n,\ell,r)&=&\notag\\
&=&\frac{1}{n_{r}!}\frac{1}{(2\kappa _{n}r)^{\ell+1}}e^{\kappa
_{n}r}\,P_{2}(n_{r},2\ell+1,2\kappa _{n}r)\,\,\,\,\,\,\,\,\,\,\,\,\,\notag \\
&&\notag\\
&+&(2\kappa _{n}r)^{\ell}e^{-\kappa
_{n}r} L_{1}(n_{r},2\ell+1,2\kappa _{n}r){Ei}(1,-2\kappa _{n}r)\,\, .\notag\\
&&\notag\\
\label{eq-R2D2}
\end{align}

\end{mdframed}

\noindent where $L_{1}(n_r,2\ell+1,2\kappa _{n}r)=L_{n_r}^{2\ell+1}(2\kappa _{n}r)$ are
associated Laguerre polynomials; $\kappa _{n}=\kappa _{0}/n$, $n=1,2,\cdots $%
, $\ell=0,1,\cdots ,n-1$; and the polynomials $P_{2}(n_{r},m,x)$ are defined in
equation (\ref{eq-P2}), and simplified in (\ref{eq-Psimple}).

\vspace{0.2cm}

The conventional second solution is commonly represented by a logarithmic term $\ln{(r)}$ times the first solution together with an infinite Laurent series. Such a series can be found by expanding the exponential integral in equation (\ref{eq-R2D2}) into powers of $r$.  The exponential integral then contributes a $\ln{(r)}$ term. Since the factor in front of the exponential function $Ei$ is a first solution, one can replace $\ln{(r)}$ by $\ln{(cr)}$, with $c$ a constant, and still have a
solution to the SE. In this way, units of the radius can be restored, and other constant
terms in the expansion of the exponential integral can be removed.

\section{Examples of second solutions}

To show the simplicity of our expressions for the second solutions to the Coulombic radial Schr\"odinger 
equation, and for reference, we give below some examples of $R_2(n,\ell,r)$.
In these examples, the radial coordinate $r$  is measured in units of the Bohr radius (corrected for
reduced mass) divided by the atomic number $Z$.

\[
R_{2}\left( 1,0,r\right) =\frac{1}{2r}e^{r}\,+e^{-r}
\int_{-2r}^{\infty }\frac{e^{-s}}{s}\,ds
\]

\[
R_{2}(2,0,r)=\frac{1}{r}\left( 1-r\right) e^{\frac{1}{2}r}+\left( 2-  r\right) e^{-\frac{1}{%
2}r} \int_{-r}^{\infty }\frac{e^{-s}}{s}\,ds
\]

\[
R_{2}\left( 2,1,r\right) =\frac{1}{r^{2}}e^{\frac{1}{2}r}\,(2+r+r^{2})+re^{-\frac{1}{2}r}%
\int_{-r}^{\infty }\frac{e^{-s}}{s}\,ds
\]

{\footnotesize

\[
R_{2}\left( 3,0,r\right) =\left( \frac{3}{2r}-\frac{5}{2}+\frac{1}{3}%
r\right) e^{\frac{1}{3}r}+\left( 3  -2r+\frac{2}{9}r^{2}\right)
e^{-\frac{1}{3}r}\int_{-2r/3}^{\infty }\frac{e^{-s}}{s}ds
\]

\[
R_{2}(3,1,r)=\left( \frac{9}{2r^{2}}+\frac{3}{r}+3-\frac{2}{3}r\right) e^{%
\frac{1}{3}r}+\left( \frac{8}{3}r-\frac{4}{9}r^{2}\right) e^{-\frac{1}{3}%
r}\int_{-2r/3}^{\infty }\frac{e^{-s}}{s}\,ds
\]

\[
R_{2}\left( 3,2,r\right) =R_{20}\left( 3,r\right) =\left( \frac{81}{r^{3}}+%
\frac{27}{2r^{2}}+\frac{3}{r}+1+  \frac{2}{3}r\right) e^{\frac{1}{3%
}r}+\frac{4}{9}r^{2}e^{-\frac{1}{3}r}\int_{-2r/3}^{\infty }\frac{e^{-s}}{s}ds
\]

\[
R_{2}\left( 4,0,r\right) =\left( \frac{2}{r}-\frac{13}{3}+\frac{22}{24}%
r  -  \frac{1}{24}r^{2}\right) e^{\frac{1}{4}r}+\left(
4-3r+\frac{1}{2}  r^{2}-\frac{1}{48}r^{3}\right) e^{-\frac{1}{4}%
r}\int_{-r/2}^{\infty }\frac{e^{-s}}{s}ds
\]

\[
R_{2}\left( 4,1,r\right) =\left( \frac{8}{r^{2}}+\frac{6}{r}+6-\frac{9}{4}%
r+  \frac{1}{8}r^{2}\right) e^{\frac{1}{4}r}+\left( 5r-\frac{5}{4}%
r^{2}+\frac{1}{16}  r^{3}\right) e^{-\frac{1}{4}%
r}\int_{-r/2}^{\infty }\frac{e^{-s}}{s}ds
\]

\[
R_{2}\left( 4,2,r\right) =\left( \frac{192}{r^{3}}+\frac{48}{r^{2}}+\frac{12%
}{r}+4+  \frac{5}{2}r-\frac{1}{4}r^{2}\right) e^{\frac{1}{4}%
r}+\left( \frac{3}{2}r^{2}-\frac{1}{8}r^{3}  \right) e^{-\frac{1}{4%
}r}\int_{-r/2}^{\infty }\frac{e^{-s}}{s}ds
\]

\[
R_{2}\left( 4,3,r\right) =R_{20}\left( 4,r\right) =\left( \frac{11\,520}{%
r^{4}}+\frac{960}{r^{3}}+\frac{96}{r^{2}}+\frac{12}{r}  +2+\frac{1%
}{2}r+\frac{1}{4}r^{2}\right) e^{\frac{1}{4}r}+\frac{1}{8} 
r^{3}e^{-\frac{1}{4}r}\int_{-r/2}^{\infty }\frac{e^{-s}}{s}ds
\]
}

\section{Conclusions}

Even though both independent solutions to the quantum Coulomb problem have been extensively
studied, the existence of explicit closed-form second solutions seems not 
to have been noticed. In a commonly-used technique,
second solutions for degenerate differential equations, a logarithm of 
the independent variable times the first solution is added to an infinite series
of terms whose coefficients are calculated. For the Coulomb case, the logarithm can be justified by 
a limiting process leading to an
expression with a term that differentiates the first solution with respect to the angular momentum quantum number $\ell$.  Because
 $r^{\ell}$ is a factor in the first solution, a $ln(r)$ times the
first solution occurs. The rest 
of the terms in the second solution become an infinite Laurent series in the radius. However,
an infinite series representation of the second (irregular) solutions can be avoided by
the method employed in this paper. Rather than an infinite series, the results
here give irregular solutions as two terms, the first having only a polynomial of degree
$(n+\ell-1)$ times
$r^{-\ell-1}exp(Zr/(na))$, and a second term expressed by an associated
Laguerre polynomial of degree $(n-l-1)$ times the factor $r^{\ell}exp(-Zr/(na))$ times the exponential integral $\mbox{Ei}(1,-2Zr/(na))$.

\section{Acknowledgements} 

This work was motivated by Prof. C.P. Gandhi, 
Head of the Department of Mathematics, Faculty of Science, Rayat Bahra University, 
Mohali, Panjab, India, whose own work on finding second solutions to differential equations is extensive.


\bibliographystyle{plain}

\end{document}